\documentclass[twocolumn,showpacs,preprintnumbers,amsmath,amssymb]{revtex4}

\usepackage{graphicx}
\usepackage{dcolumn}

\begin{document}


\title{ Measurement of a Complete Set of Nuclides, Cross-sections and
Kinetic Energies in Spallation   of $^{238}$U 1A GeV with Protons.}


\author{P.~Armbruster$^{1}$,
J.~Benlliure$^{1,2}$,
M.~Bernas$^{3}$,
A.~Boudard$^{4}$,
E.~Casarejos$^{2}$\footnote{Present address:Centre de Recherches du
Cyclotron, UCL, B-1348 Louvain-la-Neuve, Belgium},
 S.~Czajkowski$^{5}$,
T.~Enqvist$^{1}$\footnote{Present address:  CUPP-project P. O. Box 22
FIN-86801 Pyh\"asalmi,  Finland},
S.~Leray$^{4}$,
P.~Napolitani$^{1,3}$,
J.~Pereira$^{2}$,
F.~Rejmund$^{1,3}$,
M.-V.~Ricciardi$^{1}$,
K.-H.~Schmidt$^{1}$,
C.~St\'ephan$^{3}$,
J.~Taieb$^{1,3}$\footnote{Present address: CEA/Saclay
DM2S/SERMA/LENR, 91191 Gif/Yvette CEDEX, France},
L.~Tassan-Got$^{3}$,
C.~Volant$^{4}$.}

\affiliation{$^{1}$~Gesellschaft f\"ur Schwerionenforschung,
Planckstr.~1, 64291~Darmstadt, Germany}
\affiliation{$^{2}$~Universidad de Santiago de Compostela, 15706
Santiago de Compostela, Spain}
\affiliation{$^{3}$~Institut de Physique Nucl\'eaire, BP 1,
91406 Orsay Cedex, France}
\affiliation{$^{4}$~DAPNIA/SPhN, CEA/Saclay, 91191 Gif sur Yvette Cedex,
France}
\affiliation{$^{5}$~CEN Bordeaux-Gradignan, Le Haut-Vigneau, 33175
Gradignan Cedex, France}


\begin{abstract}
Spallation residues and fission fragments from 1A GeV $^{238}$U projectiles
irradiating a liquid hydrogen target were investigated by using the FRagment
Separator at GSI for magnetic selection of reaction products including
ray-tracing, energy-loss and time-of-flight techniques. The
longitudinal-momentum spectra of identified fragments were analysed, and
evaporation
residues and fission fragments could be separated. For 1385 nuclides,
production cross-sections covering 3 orders of magnitude with a mean
accuracy of 15\%, velocities in the U-rest frame and kinetic
energies were determined. In the reaction all elements from uranium to
nitrogen were found, each with
a large number of isotopes.

\end{abstract}

\maketitle



\begin{figure*}[pt]
\begin{center}
\includegraphics[ scale = 0.69]{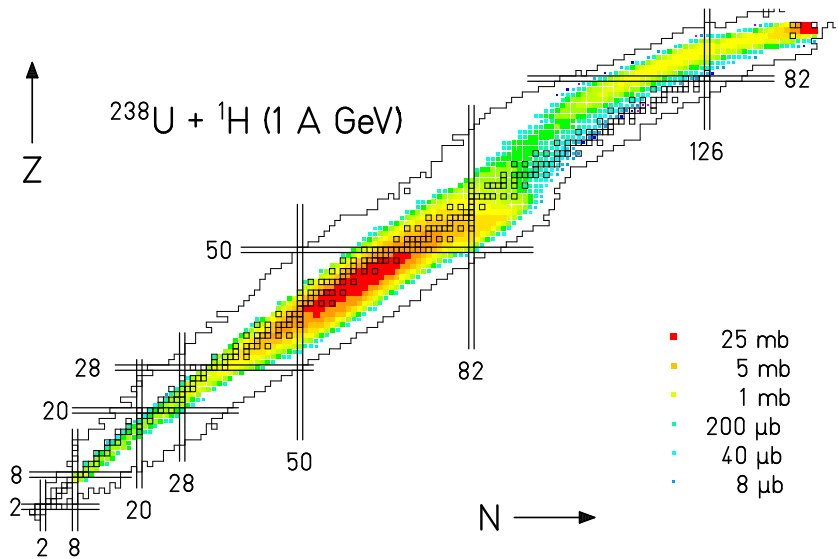}
\end{center}
\caption { The identified isotopes are shown on a chart of nuclei.
Numerical values are available on
http://www-w2k.gsi.de/kschmidt/data.htm. The logarithm of the
experimental cross-sections are indicated by a color scale.}
\label{fig:fig1}
\end{figure*}

 In view of the importance of proton-induced spallation reactions in the 1
 GeV range for future technological
 applications and the unique experimental possibilities at GSI \cite{Gei,Enq},
 Darmstadt, a program was initiated to measure isotopic
 cross-sections and kinetic energies.  ISOL-separators world-wide use the
 proton on $^{238}$U reaction since
  35 years, and future radioactive-beam facilities producing neutron-rich
  isotopes count on  it, but a solid base for the primary isotopic production
  is missing. By 1 GeV protons two main reaction channels are populated:
   Spallation Evaporation Residues (EVR) and Fission Fragments (FF).
  In the 1960's
 G. Friedlander and collaborators were the first, using radiochemical and
 surface-ionization techniques, who studied the reactions
 p + U and p + W at various
   energies  \cite{Fri}. Isotopes close to the target-nuclei and of
   the alkaline  elements rubidium
  and cesium were identified and measured.
    In the 1970's systematic measurements
  on all alkalines were undertaken using on-line mass separators
  \cite{Kla,Bel}. Later investigations of FF's in 1 GeV proton reactions on
  heavy elements  using different techniques were performed
  \cite{Boc,Vai,Tit}. All results on p-induced fission were evaluated
  recently \cite{Pro}.

    A  1A GeV $^{238}$U beam from the GSI accelerator facility produced
    EVR's and FF's in a liquid H$_2$-target,
  87 mg/cm$^2$
  thick. The target was provided by DAPNIA-Saclay and IPN-Orsay \cite{Che}.
  Fully stripped residues produced in inverse kinematics and emitted into a
  small cone in forward direction were
  separated  with well controlled efficiencies within 0.3 $\mu$s
  by the high-resolution spectrometer FRS
  \cite{Gei} and its
  ToF and $\Delta$E detectors.
  The high resolving power of the FRS enabled
 to scan the  longitudinal velocity distributions of all isotopes and to
 analyse the kinematics of the reactions in the uranium rest frame. For further
 experimental information, see refs. \cite{Enq,Tai,Ber}.
  Based only on physical
  properties of the radioactive ions, the method does not depend on
 chemistry.  Primary residues are observed, as all $\beta$-decay and nearly
  all $\alpha$-decay half-lives
  are longer than the separation times. Finally, cross-sections for
  about 1400 isotopes and their kinetic energies were determined.

  In Fig. 1 we present in a proton-neutron plot cross-sections using a colour
  logarithmic scale. In this unique comprehensive ``transmutation'' of
  uranium, all elements from uranium to nitrogen, each with a large number
  of isotopes are observed.
    In a cross-section range of about 3 orders of
  magnitude 1385 isotopes are observed, as shown on the figure. The total
  cross-section of (1.97 $\pm$ 0.3) b divides in (1.53 $\pm$ 0.2) b for fission
  \cite{Ber} and (0.44 $\pm$ 0.1) for EVR's \cite{Tai}. It is of primary interest to
  understand
  the large fraction (78 $\%$) observed for nuclear fission. For elements
  beyond tungsten, Z $>$ 74, EVR's dominate over fission fragments \cite{Tai},
  whereas
  FF's identified by their kinematics of a binary break-up process populate
  all the range below tungsten down to the lightest element nitrogen.
  Data on fission fragments are published for Z = 28 to 63  \cite{Ber}.
  The missing lightest and the heaviest elements, Z = 7-31 \cite{Val} and
  Z = 64-74 \cite{Ber3}, respectively, accomplish the full distribution
  which is presented as a high-light of this letter.

 Fig. 2 a) b) show integrated distributions of cross-sections depending on atomic and
   neutron numbers. The Z-distribution of EVR's, Fig. 2a, decreases steadily
   going
   to lighter elements down to Z = 74. A shoulder is seen in the range Z =
   80 to 88.  The distributions of FF's reveal a small 5$\%$
   contribution of the  classical asymmetric low-energy fission \cite{Ber}.
   The underlying parent nuclei are relatively
   cold. They cluster around $^{233}$U, but may reach down to A$_0$ = 226
   \cite{KHS}. Having
   separated this asymmetric contribution from the total Z-distribution, a
   mean atomic number of Z = 44.9 is found. The remaining distribution is
   surprisingly symmetric and has a standard deviation of 6.4 charge units.
   A contribution from lighter fissioning parent nuclei, which
   should increase the cross section of the lighter part of the distribution
   is barely visible in the range Z = 20 to 30. At very large
   mass-asymmetries of the FF's, Z$_1$/Z$_2$
   $<$ 20/70, the cross sections pass a minimum and increase slightly for most
   extreme asymmetries. This was seen before \cite{Sar} and is explained by
   the Businaro-Gallone mountain \cite{Bus} in the liquid-drop potential energy
   surface (LDM-PES).
\begin{figure}[h!]
\begin{center}
\includegraphics[width=\columnwidth]{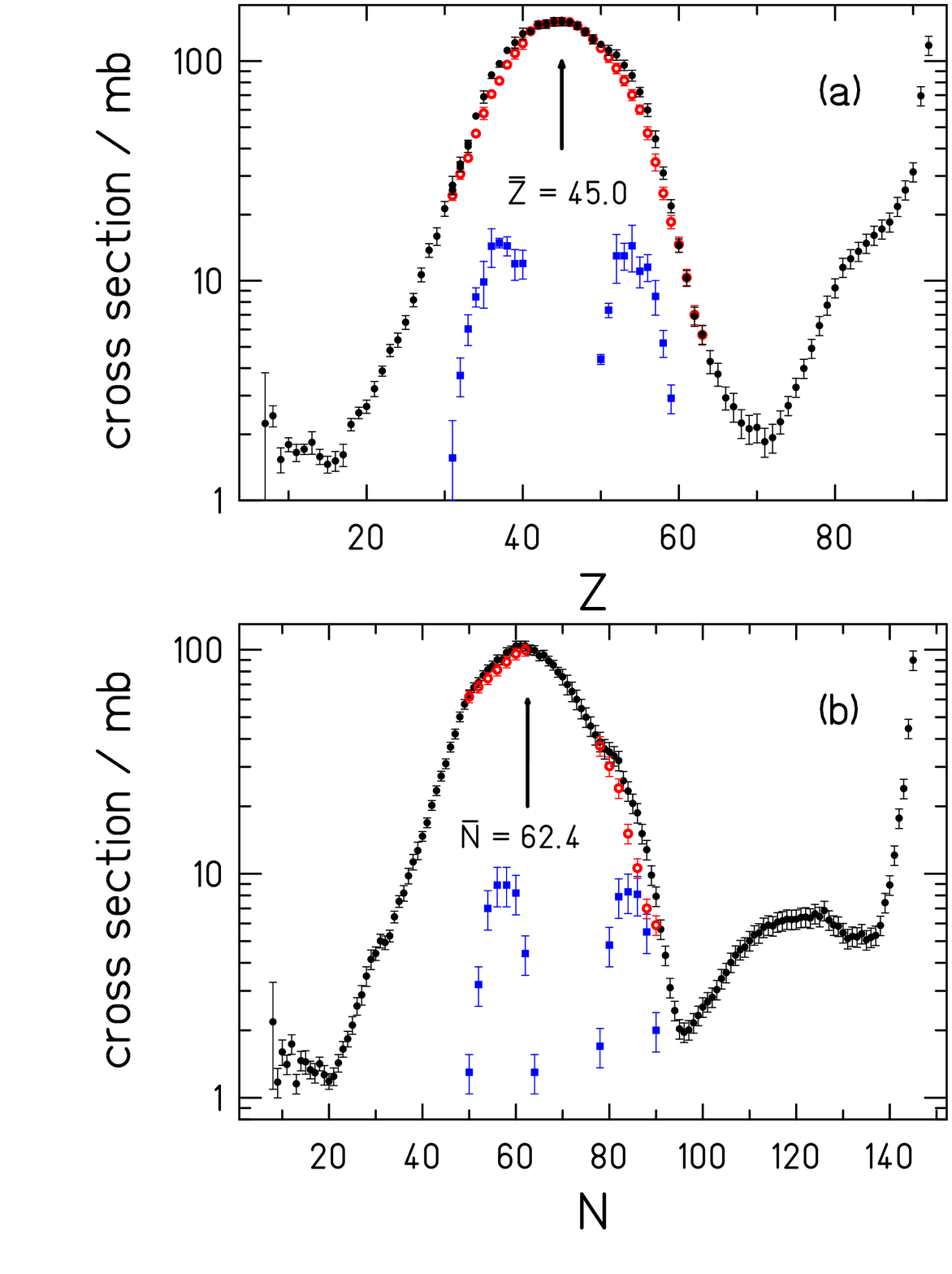}
\end{center}
\caption
{ a) Measured Z-distribution
 for all elements between Z = 7 and 92;
  b) Measured N-distribution for all neutron numbers between N = 8 and 146.
  The total cross sections (full point), cross section for low-energy
  asymmetric
  fission (blue square) and for high-energy fission (red empty point) are reported
  separately. }
\label{fig:fig2}
\end{figure}
        The N-distribution of EVR's, Fig.~2b, shows an extended plateau
     between N = 138 and N = 110 at a low level of (5.0 to 6.4) mb which
     specifies highly fissile parent nuclei (Z$^2_0$/A$_0$ $>$ 34).
     These small cross
     sections are observed for spherical nuclei around N = 126 having
   large ground-state shell-corrections and thus also increased fission barriers.
    At the excitation energies of the fissioning parent nuclei produced in
     our reaction, the higher barriers give no increased survival against
     fission.
     The  enhanced fission probabilities at N = 126 are explained by the
     low
     level densities for nuclei with spherical ground states  \cite{Jun}.
      Small cross-sections are observed as well for highly fissionable nuclei,
      being deformed  and  having much
     smaller ground-state shell-corrections  at the upper limit N = 138
     of the plateau. They are found even down at the lower limit, N = 110
     in the region of Z = 80 to 82.
     Finally,  in the range of N = 110 to 100, cross sections break down.
      At 1A GeV the
     excitation energy transferred in the reaction approaches
     its upper limit  at about
     500 MeV for mass-losses of $\Delta$A $>$ 50.
     The N-distribution of FF's, Fig. 2b, peaks at N = 62.4. The low-energy
     asymmetric fission subtracted, we obtain the high-energy symmetric
     distribution  with a mean neutron number of 61.9. Combined with the
     mean proton number 44.9 obtained from the Z-distribution a mean mass
     number of A = 106.8 is reconstructed for FF's  of high-energy
     symmetric fission.
  A small surplus of
     cross section at N = (30$\pm$6) may indicate fission from lighter parent
     nuclei down in the range of Z$_0$ = 80$\pm$4. These could be the
     asymmetric fragments of the tail of the symmetric fission channel,
     originating from fissioning parent nuclei located at
     the lower  limit  in the plateau and in its fall.

   The measured standard deviation of the high-energy symmetric Z-distribution,
   Fig. 2a, of $\sigma_Z$ =
     6.4 a.u. is related via the curvature of the LDM-PES to the excitation
     energy of the mean parent nucleus at the fission barrier
     \cite{Rus,Mul,Ben}.
     With a fission barrier of 4 MeV, an energy above ground state
     of (58 $\pm$ 10) MeV is obtained, allowing for an emission of 6 neutrons.
     Adding these neutrons emitted from the fragments to the
     neutron number of the mean pair of fragments ( 2 x 61.9 ),
       a neutron number of N$_0$ = (130$\pm$1)
      follows for the mean  parent nucleus.
       $^{220}$Th is the
     mean parent nucleus reconstructed from the isotopic distribution of FF's.

\begin{figure}[b!]
\begin{center}
\includegraphics[width=\columnwidth]{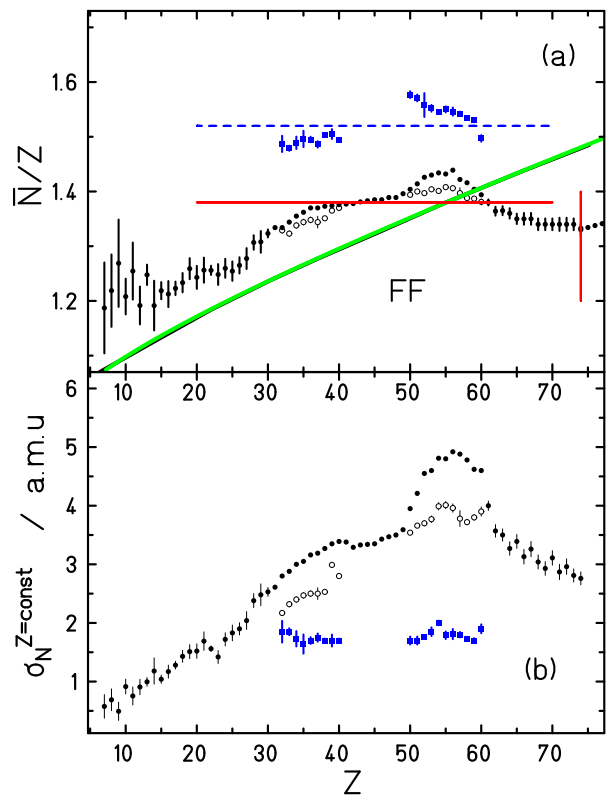}
\end{center}
\caption
{ a) The mean isotopic neutron number $\overline{N}$/Z  and
  b)  the standard deviation
$\sigma^{Z=const}_N$ are plotted as function of the atomic number.
Symbols present the different contributions (see Fig. 2). The valley of
stability (green line) and the mean N/Z ratio for low-energy asymmetric
process (dashed blue-line) and high-energy process (red line) are
indicated. The vertical arrow at Z = 74 separates fission fragments from
spallation residues. }
\label{fig:fig3}
\end{figure}

     For the complete set of FF's produced in the reaction, we obtain for
     each element the mean neutron to proton ratio $\overline{N}$/Z and the
     width of its
     isotopic distribution $\sigma_N^{Z=const.}$. These values are shown in
     Fig. 3 a), b) separated into low-energy asymmetric and high-energy
     symmetric fission. They describe the isospin dependences of the cross
     sections.
          The  $\overline{N}$/Z-ratio
      of the mean fission fragment $^{107}$Rh, 1.38, is found
     smaller than for low-energy energy fission where $\overline{N}$/Z = 1.53.

\begin{figure}[t!]
\begin{center}
\includegraphics[width=\columnwidth]{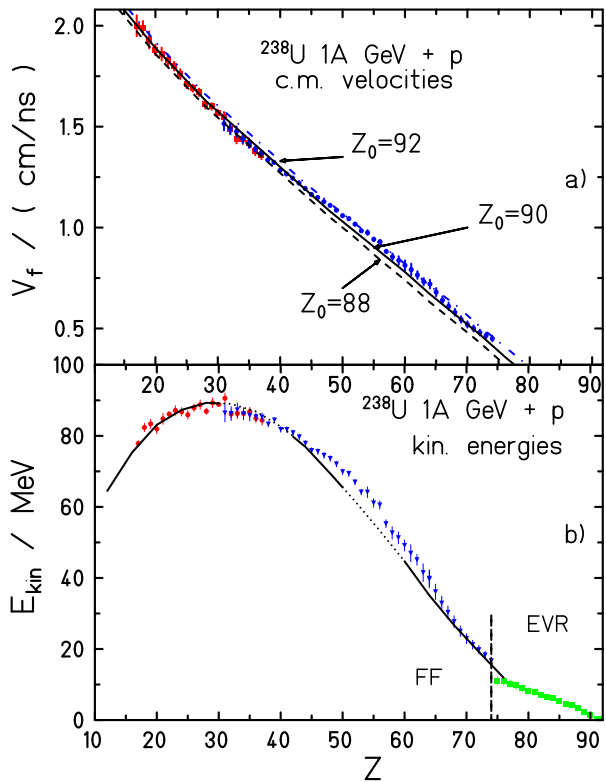}
\end{center}
\caption
{ a) The c.m. velocity of fission fragments measured
 as a function of the atomic
number. The three lines: dashed Z$_0$ = 88, full line Z$_0$ = 90, and
dashed-dotted line Z$_0$ = 92, are calculated assuming Coulomb repulsion
with a radius constant r$_0$ being kept constant and fixed
by the measured value of the velocity for symmetric fission
of $^{220}$Th  taken as normalisation.  The blue triangles and the red points
refer to ref.
 \cite{Ber} and to ref. \cite{Val} respectively.\\
 b) The kinetic energies of the isotopes as a function of the atomic
 numbers, symbols as above and green squares
 from ref. \cite{Tai}. The full line is a calculation for $^{220}$Th
 using the conditions as in the legend above.
 Asymmetric fission contributes in the regions indicated by the dotted lines.
}
\label{fig:fig4}
\end{figure}

     Fig. 3a shows in the range Z = 34 to 56
     increasing $\overline{N}$/Z-ratios for high-energy symmetric fission.
     The slope observed agrees with a
     charge-polarisation expected for a smooth LDM-PES showing no nuclear
     structure effects\cite{Arm}. A new finding is the rapid decrease of the
     mean neutron density for isotopes at higher asymmetries. Taking $^{220}$Th
     as the mean parent nucleus and the most asymmetric pair observed
     Z$_1$/Z$_2$ = 16/74, mean neutron numbers N$_1$/N$_2$ = 19/99 are
     reached summing up to 118 neutrons present in the FFs. 12 neutrons are lost
     indicating a high excitation energy of about 100 MeV, which is distributed
     in the high-energy regime between the pair of FFs proportionally to their
     masses. The heaviest elements lose up to 10 neutron, and beyond
     erbium, Z $>$ 68,  all isotopes observed are stable or proton-rich.
     Neutron-rich isotopes in the wings of the
     Z-distribution  will come with very low cross sections for the higher
     elements.
     It is the small contribution of low-energy asymmetric fission of
     (105$\pm$10)mb which stays the main source of neutron-rich
     isotopes  for elements in the range of Z = 28-64  \cite{Ber2}.

     The standard deviation   $\sigma_N^{Z=const}$ of the isotopic distribution for a
      given element
     is presented in Fig. 3b. The mean value of
    $\sigma_N^{Z=const}$ = (3.3$\pm$0.2)~a.u. compares well to
    $\sigma_N^{Z=const}$ = (3.2$\pm$ 0.7)a.u. measured in $^{238}$U
    0.75A GeV on $^{208}$Pb \cite{Sch}. For asymmetric low-energy fission,
    $\sigma_N^{Z=const}$ = (1.8$\pm$ 0.2) a.u. agrees with
    $\sigma_N^{Z=const}$ = (1.7 $\pm$ 0.05) a.u. from
   $^{238}$U 0.75A GeV on $^{208}$Pb \cite{Don}. High-energy symmetric
   fission shows a mean standard deviation wider by a factor 1.8 compared to
   asymmetric low-energy fission. The ratio of standard deviations for the two fission
   mechanisms decreases from 2.2 for barium (Z = 56) to values close to one
   for the lighter elements. This trend to a larger width of the isotopic
   distributions going to heavier elements reflects the widening of
   the LDM-PES in the (N-Z)-degree of freedom and the extended range of
   isotopes contributing to fission.
   As observed for the $\overline{N}$/Z ratio,  $\sigma_N^{Z=const}$ decreases
   for Z$>$ 56.
   The shift to smaller values for elements between Z = 58-74 is also
   visible
   in Fig. 1 showing the long plane of low cross sections coloured in green.

   Fig. 4 a) b) shows the mean velocities of FF's in the U-rest frame and
   kinetic energies
   for the elements produced as EVR's or FF's.
   The mean velocities of FF's, Fig. 4a, show a dependence which
   demonstrates the small number of elements dominating the family of
   parent nuclei in the range Z$_0$ = (88-92).
    The systematic measurement of
   kinetic energies of spallation EVR's, Fig. 4b, is a primer achieved by our
   experimental method. Their kinetic energies are very small, on the
   average about 2.9 MeV. In this energy range slowing down of heavy ions
   mainly  proceeds
   by elastic collisions.
    The maximum of kinetic energies at one third of the atomic number of
   the parent nucleus, as expected for symmetric fission of a monoisotopic
   fission source, is experimentally verified.

   The complete data-set presented is a main step forward, which  in this
   letter stands for itself. Certainly our work is not finished here.
   The tasks to be
   done in the near future are open:



\noindent  -~~  The energy
   dependence of the cross sections is hardly known. Further measurements
   at lower energies are needed for our reaction.\\
  -~~   Our data should serve as a bench mark
   for simulation codes of the complex physics of spallation reactions
   with the final goal to predict unknown
   systems.\\
  -~~   An innovative measuring technique giving more and more precise and complete
   results generates a better understanding of the underlying physics.
   Further contributions to fundamental aspects of spallation reactions
   may still emerge in a coming analysis.\\

  Financial support by EU-contract ERBCHBCT940717 for J. B., T. E. and F. R.
 and by GSI for P. N., M. V. R. and J. T. is gratefully acknowledged.
 This work was partially supported by the European Union in the
 HINDAS project (contract FIKW-CT-20000-00031),
 and by the support to the access to large facilities (contract EC-HPRI-CT-1999-00001.).

\end{document}